\documentclass{ws-ijmpa}
\usepackage {graphicx} 
\usepackage{amsmath}
\begin{document}

\markboth{Rossi Zaninetti}
{Torsion Balance}

%
\catchline{}{}{}{}{}
%
\def\aplett{Astrophys.~Lett.\,}
\def\apj{ApJ\,}
\def\apjl{ApJ\,}
\def\aap{A\&A\,}
\def\mnras{MNRAS\,}
\def\pasj{PASJ\,}
\def\solphys{Sol.~Phys.\,}
\def\aj{AJ}
\def\aaps{A\&AS}
\def\araa{ARA\&A}
\def\apjs{ApJS}
\def\cjaa{Chinese J. Astron. Astrophys.}
\def\jaa{J. Astrophys. Astr.}
\def\planss{Planet.~Space~Sci.}
\def\apss{Astrophysics and Space Science}
\def\pasp{PASP}
\def\JPG{J. Phys. G\,}
\def\POF{Physics of Fluids}
\def\physrep{Phys.~Rep.\,}
\def\nat{Nature\,}
\def\pre{Phys.~Rev.~E}
\def\pra{Phys.~Rev.~A}
\def\physa{Physica A}
\def\rpp{Rep.Prog.Phys.}
\title
{Linear and Non Linear Effects on the
Newtonian Gravitational Constant as  deduced  from the
Torsion Balance }

\author     {M. Rossi}
\address    {Dipartimento di Matematica, \\
      Universit\`a degli Studi di Torino \\
Via Carlo Alberto 10,\\ 10123 Torino, Italy}

\author{ L. Zaninetti}

\address {Dipartimento di Fisica Generale, \\
      Universit\`a degli Studi di Torino \\Via Pietro Giuria 1,\\
           I-10125 Torino, Italy
\\
phone +390116707460, 
fax   +390116699579
}

\maketitle

\begin{history}
\received{Day Month Year}
\revised{Day Month Year}
\end{history}

\begin{abstract}
The Newtonian gravitational constant has still 150  parts per
million of uncertainty. This paper examines the linear and
nonlinear equations governing 
the rotational dynamics of the torsion gra\-vi\-ta\-tional balance. 
A nonlinear effect modifying the
oscillation period of the torsion gravitational balance  is
carefully explored.


\end{abstract}

\ccode{PACS numbers: 04.80.-y ;
06.20.Jr;
 }

\section{Introduction}
After many years of measurements, begun by H. Cavendish, the
Newtonian gravitational constant value, said $G$ , is still
affected by a large error~\cite{Gillies1997}. The CODATA
recommends $G=(6.6742 \pm 0.001)\times 10^{-11}\frac { m^3} {kg
s^2}$, meaning a relative standard uncertainty of 150 parts per
million (in the following ppm)~\cite{CODATA2005}. This value has
had recent confirmation by means, on one hand, of a
super-conducting gravimeter~\cite{Baldi2005} and, on the other
hand, of a careful analysis of the possible beam balance
nonlinearity~\cite{Schlamminger2006}. The original Cavendish
method of measure, employing the torsion balance, still 
reveals  a large discrepancy
from the recommended value: about
500~ppm~\cite{Schurr_1998}~. This is probably due to imperfections
of the crystalline structure of the torsion
fibre~\cite{Bagley1997,Kuroda1995,Matsumura1998}. 
Moreover recent studies point out that the period of a torsion pendulum
might vary under disturbances of environmental noise factors, see
see~\cite{Luo2005}.
Other authors suggest 
a  possible deviation
from Newton's  law    specified as an additional contribution of Yukawa
potential type ~\cite{Kononogov2005}.
This paper first analyses the linear and non linear equations
governing the torsional balance rotational dynamics
(Section~\ref{section1}). By means of a gravitational torsion
balance, same values of $G$ are obtained and summarised
(Section~\ref{section2}). The oscillation period's variation, due
to a non linear effect, is then discussed
(Section~\ref{section3}).

\section{The basic equations}
\label{section1}
\begin{figure*}
\begin{center}
\includegraphics[width=8 cm,angle=-90]{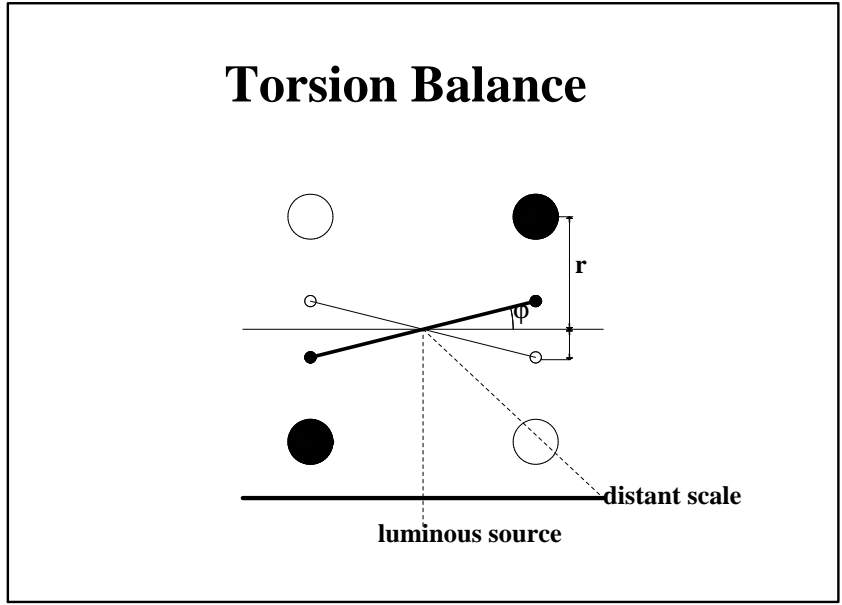}
\end {center}
\caption {Schematic view of the torsional balance}
\label{primafigura}%
\end{figure*}

\begin{figure*}
\begin{center}
\includegraphics[width=8cm,angle=-90]{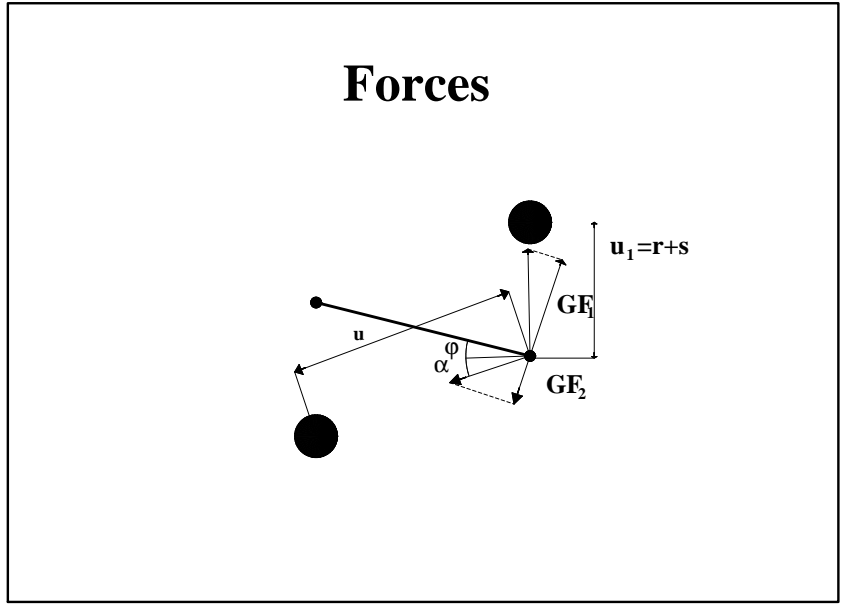}
\end {center}
\caption {Top view of the Cavendish balance}
\label{secondafigura}%
\end{figure*}

The form of Newton's  law of gravitation is
\begin{equation}  \label{newton}
F=G\frac{mM}{r^{2}}
\quad ,
\end{equation}
where  $G$ is the gravitational constant , $M$  the great mass , $m$ the
small mass and $r$ their relative distance. The Leybold  balance
represents a widespread  instrument to determine the constant $G$,
see Figure~\ref{primafigura},
and  
is constructed with the following components.
\begin{enumerate}
\item  A freely oscillating horizontal bar , of length $2d$, holding
two small lead balls of mass  $m$
as in Figure~\ref{primafigura} supported by
a torsion fibre   that has a
 torsional constant  $\tau $~.
\item  Two larger balls of mass $M$  that can be positioned next
to the small balls  as in Figure~\ref{primafigura}. The center of mass
of the two $m$ and $M$ are supposed to be all on a plane
perpendicular to the fibre. \item   A luminous source  directed
toward the center of mass of the bar  where is reflected by a
mirror.
\item   A  scale at distance {\it l}
\/where the  reflected
light beam is measured.
\end{enumerate}
Thus the equilibrium
position about which the pendulum oscillates is different
for the two positions and it is this difference
which we use to determine G. The  Figure~\ref{secondafigura}
reports  a plot of the motion.
\noindent
The  moment of inertia  of the bar,$I$  is
\begin{equation}
I\sim 2md^{2}  \label{In}
\quad.
\end{equation}
The fundamental equation of rotational dynamics
is
\begin{equation}
I\stackrel{..}{\varphi }=M_{g}+M_{v}+M_{t}
\quad ,
\label{eqcard}
\end{equation}
where
\begin{equation}
M_{v}=-\beta \stackrel{.}{\varphi }
 \label{Mv}
\quad ,
\end{equation}
\begin{equation}
M_{t}=-\tau \varphi  \label{Mt}
\quad ,
\end{equation}
here  $\beta $
is the  coefficient of viscosity of air ,
$\varphi $ the angle  between bar and bar itself
when the torque is zero.
This angle  is measured in the anti clockwise direction.
The term  $M_{g}$ represents the torque of the
gravitational forces.
From  equation~$ ({\ref{newton}}) $ ,
we obtain
\begin{equation}
M_{g}=2dGF ( \varphi )  \label{Mg}
\quad ,
\end{equation}
where  $F$ is a function of the angle $\varphi $. The law of
dependence of  $F$  with $\varphi $  is complex and will  here be
analysed.
 When the motion starts the
resulting force is
\begin{equation}  \label{F}
F( \varphi ) =F_{1}(\varphi ) -F_{2}( \varphi )
\quad ,
\end{equation}
where
\begin{equation}  \label{F1}
F_{1}( \varphi ) =mM\frac{\cos \varphi }{u_{1}^{2}}
\quad ,
\end{equation}
\begin{equation}  \label{F2}
F_{2}( \varphi ) =mM\frac{\cos ( \frac{\pi }{2}-( \varphi +\alpha
) ) }{u^{2}}=mM\frac{\sin ( \varphi +\alpha ) }{u^{2}}
\quad ,
\end{equation}
now $\alpha :=\arcsin ( \frac{r-s}{u}) $ (where $r$ e $s$ are
defined as in Figure~\ref{primafigura}).
With our data , see
Section~\ref{data} the maximum angular excursion
 of the angle
$\varphi $ is
\begin{equation}
\Delta  =\arcsin ( \frac{x_{\max }-x_{\min }}{2l})
\quad.
\end{equation}
The angle  $\varphi$ has a low values, see Table~\ref{parameters},
and a Taylor series expansion that keep terms to order $\varphi^2$
will be adopted. This means to forget quantities less than
 3~$10^{-5}$.
The series representation gives
\begin{equation}
u_{1}=\sqrt{( r+s) ^{2}+( d-d\cos \varphi ) ^{2}}\sim
r( 1+\varphi ( \frac{2d}{r}+\frac{d^{2}}{r^{2}}\varphi +\frac{d^{2}%
}{4r^{2}}\varphi ^{3}) ) ^{\frac{1}{2}} \quad.
\end{equation}
Developing the last term with a  Maclaurin  series  we obtain
\begin{equation}
u_{1}\sim r( 1-\frac{\varphi }{2}( \frac{2d}{r}+\frac{d^{2}}{r^{2}}%
\varphi ) -\frac{d^{2}}{2r^{2}}\varphi ^{2}+O( \varphi ^{3}) )
\sim r+d\varphi \quad.
\end{equation}
As a consequence
\begin{equation}
 F_{1}\sim mM\frac{1-\frac{\varphi ^{2}}{2}}{( r+d\varphi ) ^{2}}
 \quad ,
\end{equation}
and expanding the denominator we obtain
\begin{equation}
\left(r+d\varphi \right)^{-2}=({r}^{-2}-2\,{\frac{d}{{r}^{3}}}
\varphi+3\,{\frac{{d}^{2}}{{r}^{4}}}{\varphi}^{2}+O
\left( {\varphi}^{3} \right) ) \quad ,
\end{equation}
%
%
that means
\begin{equation}
F_{1}\sim \frac{mM}{r^{2}}( 1-\frac{2d}{r}\varphi +( \frac{3d^{2}}{%
r^{2}}-\frac{1}{2}) \varphi ^{2})  \label{svF1} \quad.
\end{equation}
An expression for $F_{2}$ can be obtained from equation~$( {\ref{F2}}) $
\begin{equation}
F_{2}=mM\frac{\sin \varphi \cos \alpha +\cos \varphi \sin \alpha }{u^{2}}%
\sim mM\frac{r+d\varphi -\frac{r}{2}\varphi ^{2}}{(
d^{2}+r^{2}-2rd\varphi -d^{2}\varphi ^{2}) ^{\frac{3}{2}}} \quad.
\end{equation}
On Taylor expanding the denominator
\begin{equation}
( 4d^{2}+r^{2}-2rd\varphi -d^{2}\varphi ^{2}) ^{-\frac{3}{2}%
}=( 4d^{2}+r^{2}) ^{-\frac{3}{2}}( 1+\frac{3dr}{4d^{2}+r^{2}}%
\varphi +\frac{3d^{2}( 4d^{2}+5r^{2}) }{2( 4d^{2}+r^{2})
^{2}}\varphi ^{2}+O( \varphi ^{3}) ) \quad ,
\end{equation}
and therefore
\begin{equation}
F_{2}\sim \frac{r+\frac{4d( d^{2}+r^{2}) }{4d^{2}+r^{2}}\varphi +%
\frac{r( 20d^{4}+13d^{2}r^{2}-r^{4}) }{2(
4d^{2}+r^{2}) ^{2}}\varphi ^{2}}{( 4d^{2}+r^{2}) ^{\frac{3}{%
2}}}  \label{svF2} \quad.
\end{equation}
Now  $F$ from equation~$( {\ref{F}}) $ can be expressed like a
Taylor expansion truncated
 at  $O( \varphi ^{3}) $
\begin{equation}
F\sim A_{0}+A_{1}\varphi +A_{2}\varphi ^{2}  \label{svF} \quad ,
\end{equation}
where
\begin{equation}
A_{0}:=mM( \frac{1}{r^{2}}-\frac{r}{( 4d^{2}+r^{2}) ^{\frac{3%
}{2}}})  \label{A0} \quad ,
\end{equation}
\begin{equation}
A_{1}=-2mM( \frac{d}{r^{3}}+\frac{2d( d^{2}+r^{2}) }{(
4d^{2}+r^{2}) ^{\frac{5}{2}}})  \label{A1} \quad ,
\end{equation}
\begin{equation}
A_{2}= 3\,{\frac{mM{d}^{2}}{{r}^{4}}}-1/2\,
{\frac{mM}{{r}^{2}}}+1/2\,{\frac {mMr \left(
4\,{d}^{4}-10\,{r}^{2}{d}^{2}+{r}^{4} \right)}
{\left( 4\,{d}^{2}+{r}^{2} \right) ^{7/2}}}
\quad.
\end{equation}
Three methods that allow to obtain an expression for $G$ in
terms of measurable quantities are now  introduced. Further on the well
known formula for G extracted from the Leybold manual is reviewed.

\subsection{Averaged G}
Let $F( \varphi ) =\overline{F}$ for all the experience ; in first
approximation we may assume that
 $F$ is given by equation~$( {\ref{svF}}) $
to first order
 \begin{equation}
\overline{F}=A_{0}+A_{1}\overline{\varphi }  \label{Fmedio} \quad
,
\end{equation}
where $\overline{\varphi }$ is the average of the values that
$\varphi $ assumes between  the first position , $P_{1}$,
 and the last
position  ,$P_{\infty }$,
 of the balance ;
 $A_{0},A_{1}$ are given
by equations~( {\ref{A0}}) and ({\ref{A1}})~.
 The differential
equation that describes the motion is
\begin{equation}
I\stackrel{..}{\varphi }+\beta \stackrel{.}{\varphi }+\tau \varphi =2dG%
\overline{F}  \label{eqmoto1} \quad ,
\end{equation}
and it's  solution  is
\begin{equation}
\varphi ( t) =ce^{-\delta t}\cos ( \omega t+\phi ) +%
\frac{2dG\overline{F}}{\tau }  \label{traiett1} \quad ,
\end{equation}
where  $c$ represents the \textit{amplitude} and
\begin{equation}
\delta :=-\frac{\beta }{2I} \label{delta} \quad ,
\end{equation}
\begin{equation}
\omega :=\frac{\sqrt{4I\tau -\beta ^{2}}}{2I} \quad.
\end{equation}
The angle  $\varphi _{\infty }$, that represents the bar position
at the end of the phenomena  $P_{\infty }$ can be determined as
follows
\begin{equation}
\varphi _{\infty }\sim \frac{x_{\infty }-x^{o}}{2l}=\frac{x_{\infty }-x_{1}}{%
4l} \quad ,  \label{finf}
\end{equation}
and should be the same as predicted by the theory
\begin{equation}
\lim_{t\rightarrow +\infty }\varphi ( t) =\frac{2dG\overline{F}}{%
\tau } \quad ;
\end{equation}
  therefore
\begin{equation}
G=\frac{\tau \varphi _{\infty }}{2d\overline{F}}  \label{Gprimo}
\quad.
\end{equation}
In order to continue a value for  $\tau $ should be derived. This
can be obtained from the period of oscillation of the bar
\begin{equation}
T=\frac{2\pi }{\omega }=\frac{4\pi I}{\sqrt{4I\tau -\beta ^{2}}}
\label{T} \quad.
\end{equation}
We continue by identifying $T$ with the empirical value
$\overline{T}$. We continue on assuming that $\frac{\beta
^{2}}{4I}$ is small ; therefore from equations $( {\ref{In}} ) $,
 $( {\ref{Gprimo}}) $ and  $( {\ref{Fmedio}}) $, the following is obtained
 \begin{equation}
G=\frac{2\pi ^{2}I\varphi _{\infty }}{d( A_{0}+A_{1}\overline{\varphi }%
) {\overline{T}}^2} \label{Gnum1} \quad.
\end{equation}

\subsection{G with air viscosity}

From formula~(\ref{Mv})
 is possible to deduce the viscosity of the
air once the coefficient of damping
$\delta$ is known, see
Section~\ref{data} on data analysis.
 From formula~(\ref{Gprimo})
and (\ref{delta})   we should add to the value of
  $G$ reported
in equation~(\ref{Gnum1})
\begin{equation}
 G_{\beta }:=\frac{\beta ^{2}\varphi _{\infty
}}{8Id\overline{F}}  \label{Gb} \quad ,
\end{equation}
obtaining
\begin{equation}
G=\frac{2\pi ^{2}I\varphi _{\infty }}{d( A_{0}+A_{1}\overline{\varphi }%
) {\overline{T}}^2}+\frac{\beta ^{2}\varphi _{\infty }}{8Id(
A_{0}+A_{1}\overline{\varphi }) } \quad.
 \label{Gnum1b}
\end{equation}

\subsection{G to the first order}

Let assume that  $F( \varphi ) $ is not constant, we can assume at the order
 $O( \varphi ^{2})$ with the aid of
formula~(\ref{svF})
\begin {equation}
F\sim A_{0}+A_{1}\varphi \quad ,
\end{equation}
where $A_{0}$ e $A_{1}$ are defined in equations~$( {\ref{A0}}%
) $ and $( {\ref{A1}}) $ respectively. In this case
 the law of
motion is still equation $( {\ref{eqcard}}) $
 \begin{equation}
I\stackrel{..}{\varphi }+\beta \stackrel{.}{\varphi }+( \tau
-2dGA_{1}) \varphi =2dGA_{0}  \label{eqmoto2} \quad ,
\end{equation} and the solution is
\begin{equation}
\varphi ( t) =ce^{-\delta t}\cos ( \omega ^{\prime }t+\phi )
+\frac{2dGA_{0}}{\tau -2dGA_{1}}  \label{traiett2} \quad ,
\end{equation}
where the angular velocity  $\omega ^{\prime }$ has now the
following expression
\begin{equation}
\omega ^{\prime }:=\frac{\sqrt{4I( \tau -2dGA_{1}) -\beta ^{2}}}{%
2I} \quad.
\end{equation}
As a consequence
\begin{equation}
\tau =\frac{4\pi ^{2}I}{\overline{T}^{2}}+2dGA_{1}+\frac{\beta
^{2}}{4I} \label{tors2} \quad ,
\end{equation}
\begin{equation}
\varphi _{\infty }=\lim_{t\rightarrow +\infty }\varphi ( t) =%
\frac{2dGA_{0}}{\tau -2dGA_{1}} \quad ,
\end{equation}
and therefore
\begin{equation}
G=\frac{\tau \varphi _{\infty }}{2d( A_{0}+A_{1}\varphi _{\infty
}) } \quad.
\end{equation}
Once equation~$( {\ref{tors2}}) $ is substituted in this
relationship  we obtain
\begin{equation}
G=\frac{2\pi ^{2}I\varphi _{\infty
}}{dA_{0}{\overline{T}}^2}+\frac{\beta ^{2}\varphi _{\infty
}}{8dIA_{0}} \quad.
  \label{Gnum3}
\end{equation}
\subsection{G from Leybold manual}

The deduction of G through  the Leybold torsional balance is widely
known , see \cite{leybold}. We simply report the final expression
\begin{equation}
G ={\pi^2 b^2 d \Delta S  \over{M T^2 l}}\times (1 +\beta) \quad,
\label{Gleybold}
\end{equation}
where
\begin{equation}
\beta  ={\displaystyle   b^3 \over \displaystyle {(b^2 +4d^2)
\sqrt{b^2 +4d^2}}} \quad.
\end{equation}
The meaning of the symbols is
\begin{itemize}
\item{$b$:}Distance between center of the great mass and small mass
\item{$\Delta$ S:}Total deflection of the light spot
 \item{$d$:}The length of the lever arm
 \item{$l$:}Distance between mirror and screen
\item{$M$:} Great mass \item{T} Period of the oscillations
\end{itemize}

\section{Analysis of the data}
\label{section2}
\label{data}
 The physical parameters as well their uncertainties
are reported in Table~\ref{parameters}.
\begin{table}
\caption{Parameters of the torsion balance. }
\label{parameters}
\begin{tabular}{lcl}
  \hline \hline
  parameter & value &unit    \\ \hline
  $M$         & (1.5 $\pm$ $10^{-3}$)                 & $Kg$ \\
  $m$         & (1.5$\cdot10^{-2}$ $\pm$ $10^{-3}$)   & $Kg$ \\
  $r$         & (4.65$\cdot10^{-2}$ $\pm$ $10^{-3}$)   & $m$ \\
  $d$        & (5.0$\cdot10^{-2}$ $\pm$ $10^{-3}$)   & $m$ \\
  $l$        & (5.475  $\pm$ $10^{-3}$)   & $m$ \\
  $\varphi$ &(-6.715$\cdot10^{-3}$ $\pm$ 1.3$\cdot10^{-5}$)& $rad$ \\
  $\beta$     &( 1.432$\cdot10^{-7}$ $\pm$ 1.11$\cdot10^{-8}$)&
  $\frac{kg\cdot  m^{2}}{s}$\\
   \hline\hline
\end{tabular}
\end{table}
The data were analysed through the following fitting function
\begin{equation}
y(t) = A_0 +A_1 cos ( \frac{2 \pi t} {T}) \exp (-\frac{t}{\tau})
\quad.\label{fit}
\end{equation}
 The data has been  processed through the
Levenberg--Marquardt  method ( subroutine MRQMIN in~\cite{press})
in order to find the parameters  $A_0$,$A_1$ ,$T$ and $\tau$. The
results are reported in Table~\ref{data_fit} together with the
derived quantities.
\begin{table}
\caption{Parameters of the nonlinear fit. }
\label{data_fit}
\begin{tabular}{lcl}
  \hline \hline
  parameter & value &unit    \\ \hline
  $A_0$         & (8.188$\cdot10^{-2}$ $\pm$1.11$\cdot 10^{-4}$)& $m$ \\
  $A_1$         & ( 0.1470 $\pm$ $2.85\cdot10^{-4}$)    &$ m$ \\
  $T$           & (552.98 $\pm$ 0.16)   & $s$ \\
  $\tau$        & (1047.0 $\pm$ 4.8)    & $s$ \\
    \hline\hline
\end{tabular}
\end{table}

The value of $G$  can be derived coupling the basic parameters of
the torsion balance , see Table~\ref{parameters}, and the measured
parameters of the damped oscillations , see Table~\ref{data_fit}.
Table~\ref{G_values} reports the four
values of $G$ here
considered with the uncertainties
expressed in absolute value and
in ppm ;
 the precision of the measure in
  respect of the so called
"true" value  is also reported.
A considerable source of error is the
uncertainty in the
determination of the
span between the two spheres
that in our  case is $\approx 10^{-3} m $ .
Adopting  a rotating gauge
method~\cite{Luo2001}
the uncertainty in the
determination of the
span between the two spheres is $\approx 0.5 \times 10^{-6} m $;
this is the way to lower the uncertainty in Table~\ref{G_values}.
\begin{table}
\caption{Values of $G$ }
\label{G_values}
\begin{tabular}{lllll}
  \hline \hline
 method    & equation    & value & uncertainty [ppm]& accuracy $\%$]         \\
$G$ averaged             &(\ref{Gnum1})   & (6.67 $\pm$  0.34)$\cdot10^{-11}\frac{m^{3}}{kg\cdot s^{2}}$ &  52433   & 0.0161    \\
$G$  with air viscosity &(\ref{Gnum1b}) & (6.72 $\pm$  0.35)$\cdot10^{-11}\frac{m^{3}}{kg\cdot s^{2}}$ &  52433   & 0.69         \\
$G$ to the first order  &(\ref{Gnum3})   & (6.80 $\pm$  0.34)$\cdot10^{-11}\frac{m^{3}}{kg\cdot s^{2}}$ &  49989   & 1.92         \\
$G$ from Leybold manual&(\ref{Gleybold})&  (6.71 $\pm$  0.33) $\cdot10^{-11}\frac{m^{3}}{kg\cdot s^{2}}$&   49600   & 0.64        \\
    \hline\hline
\end{tabular}
\end{table}

\section{Non linear effects in the vacuum}
\label{section3}
By starting from the equation of  rotational dynamics up
to the second order
\begin{equation}
I\stackrel{..}{\varphi }+\beta \stackrel{.}{\varphi }+( \tau
-2dGA_{1}) \varphi -2dGA_{0}=2dGA_{2}\varphi ^{2}
\quad ,
\end{equation}
 the case of  $\beta$=0 is analysed ,
\begin{equation}
I\stackrel{..}{\varphi }+( \tau
-2dGA_{1}) \varphi -2dGA_{0}=2dGA_{2}\varphi ^{2}
\quad  ,
\end{equation}
that  corresponds to perform the experiment in the vacuum.
On dropping the constant term and dividing by $I$ we obtain
\begin{equation}
\stackrel{..}{\varphi }+\frac {( \tau
-2dGA_{1})}{I} \varphi =\frac{2dGA_{2}}{I}\varphi ^{2}
\quad.
\end{equation}
On imposing
\begin {equation}
\omega_0 ^2 =\frac {( \tau
-2dGA_{1})}{I}
\quad ,
\end {equation}
the  nonlinear ordinary differential equation , in the
following ODE ,  has  the form
\begin{equation}
\stackrel{..}{\varphi }+\omega_0^2 \varphi =\frac{2dGA_{2}}{I}\varphi ^{2}
\quad.
\end{equation}
On adopting the transformation  $T = t *\omega_0$
the nonlinear  ODE  is
\begin{equation}
\stackrel{..}{\varphi }+ \varphi -  \epsilon  \varphi^2 =0
\quad ,
\label{ODEnostra}
\end{equation}
where
\begin{equation}
 \epsilon = \frac{2dGA_{2}} {( \tau
-2dGA_{1})}
\quad.
\end{equation}
The solution of equation~(\ref{ODEnostra})
is reported in the Appendix~\ref{appendix_a}
and in our case  $\epsilon$=0.0187.
We  now  have a measured period ,$T_{MS}$,
that is equalised to the  non linear value , $T_{NL}$~.
The period of the linear case , $T_L$ ,
can be written as
\begin{equation}
T_{NL} = T_{MS} = 1.00014   T_L
\quad ,
\end{equation}
and therefore
\begin{equation}
T_{L} =  \frac {T_{MS}} {1.00014}
\quad.
\end{equation}
In the various formulae of $G$ without damping , for example
equations~(\ref{Gnum1}) and~(\ref{Gleybold}),
the 
periods $T_L$ and $T_{MS}$ are raised to the square
\begin{equation}
T_{L}^2 =  \frac {T_{MS}^2} {1.00029}
\quad ,
\end{equation}
and in  the denominator,  making the
 non linear $G_{NL}$   greater
than  the linear $G_L$
\begin{equation}
G_{NL} = 1.00029 \, G_{L}
\quad.
\end {equation}
The value  of this correction , $\delta G$,
can be evaluated
as a difference between 1
and the multiplicative factor of   $G_L$
\begin{equation}
\delta G =(1.00029-1) \times 6.6742\times 10^{-11} m^3 kg^{-1} s^{-2} =
0.19\times 10^{-13}~m^3 kg^{-1} s^{-2}
\quad.
\end{equation}
The official error on $G$ is
 $0.1\times10^{-13}~m^3 kg^{-1} s^{-2}$
and therefore the nonlinear correction can be expressed as
the double of the official error on $G$.
\appendix

\section{The eardrum equation}
\label{appendix_a}
The  equation
\begin{equation}
\ddot{x}+ x +  \epsilon  x^2 =0
\quad ,
\label{eardrum}
\end{equation}
is well known under the name "eardrum equation".
It can  be solved , see~\cite{Enns2002},  transforming it  in
\begin{equation}
{\Omega}^{2}{\frac {d^{2}}{d{T}^{2}}}X \left( T \right) +X \left( T
 \right) +\epsilon\, \left( X \left( T \right)  \right) ^{2}=0
\quad ,
\end{equation}
and adopting the method of  Poisson that imposes the following solution
to  $X$
\begin{equation}
{\it x} \left( T \right)   ={\it x_0} \left( T \right) +
{\it x_1} \left( T \right) \epsilon+{\it x_2}
 \left( T \right) {\epsilon}^{2}
\quad ,
\end{equation}
and to  $\Omega$
\begin{equation}
\Omega = 1+{\it \omega_1}\,\epsilon+{\it \omega_2}\,{\epsilon}^{2}
\quad.
\end{equation}
The computer algebra system (CAS) gives
\begin{equation}
\omega_1=0   \quad \omega_2 := -5/12
\quad.
\end{equation}
\section* {Acknowledgements}
We thank Richard Enns  who   has provided us the Maple
routine Example 04-S08  extracted from~\cite{Enns2002}.

\bibliography{biblio}
\end{document}